\documentclass{IEEEtran}
\usepackage{graphicx}
\usepackage{amsmath}
\usepackage{amssymb}
\usepackage{tikz}

\usepackage{eso-pic}
\AddToShipoutPicture*{\scriptsize\sffamily\raisebox{0.6cm}{\hspace{1.65cm}\fbox{\parbox{\textwidth}{\copyright~2017 IEEE. Personal use of this material is permitted. Permission from IEEE must be obtained for all other uses, in any current or future media, including reprinting/republishing this material for advertising or promotional purposes, creating new collective works, for resale or redistribution to servers or lists, or reuse of any copyrighted component of this work in other works.}}}}

\begin{document}
\title{Multiple Right-Hand Side Techniques in Semi-Explicit Time Integration Methods for Transient Eddy Current Problems}
\author{\IEEEauthorblockN{Jennifer Dutin\'{e}\IEEEauthorrefmark{1},
Markus Clemens\IEEEauthorrefmark{1}, and
Sebastian Sch\"{o}ps\IEEEauthorrefmark{2}}\\
\IEEEauthorblockA{\IEEEauthorrefmark{1}Chair of Electromagnetic Theory,
University of Wuppertal, 42119 Wuppertal, Germany}\\
\IEEEauthorblockA{\IEEEauthorrefmark{2}Graduate School CE, Technische Universit\"{a}t Darmstadt, 64293 Darmstadt, Germany}
\thanks{ 
Corresponding author: J.  Dutin\'{e} (email: dutine@uni-wuppertal.de)}}

\maketitle

\begin{abstract}
The spatially discretized magnetic vector potential formulation of magnetoquasistatic field problems is transformed from an infinitely stiff differential algebraic equation system into a finitely stiff ordinary differential equation (ODE) system by application of a generalized Schur complement for nonconducting parts. The ODE can be integrated in time using explicit time integration schemes, e.g. the explicit Euler method. This requires the repeated evaluation of a pseudo-inverse of the discrete curl-curl matrix in nonconducting material  by the preconditioned conjugate gradient (PCG) method which forms a multiple right-hand side problem. The subspace projection extrapolation method and proper orthogonal decomposition are compared for the computation of suitable start vectors in each time step for the PCG method which reduce the number of iterations and the overall computational costs.
\end{abstract}



\section{Introduction}
\IEEEPARstart{I}{n} the design process of transformers, electric machines, etc., simulations of magnetoquasistatic field problems are an important tool. In particular in multi-query scenarios, as needed e.g. in the case of uncertainty quantification or optimization, using efficient and fast algorithms is important.  

The spatial discretization of the magnetic vector potential formulation of eddy current problems yields an infinitely stiff differential-algebraic equation system of index 1 (DAE). It can only be integrated in time using implicit time integration schemes, as e.g. the implicit Euler method, or singly diagonal implicit Runge-Kutta schemes~\cite{Hairer,Nicolet}.  Due to the nonlinear B-H-characteristic of ferromagnetic materials large nonlinear equation system have to be linearized, e.g. by the Newton-Raphson method, and resolved in every implicit time step. At least one Newton-Raphson iteration is required per time step. The Jacobian and the stiffness matrix have to be updated in every iteration.

A linearization within each time step is avoided if explicit time integration methods are used. First approaches for this were published in~\cite{Yioultsis} and~\cite{Ausserhofer}, where different methods are used in the conductive and nonconductive regions respectively. In~\cite{Yioultsis}, the Finite Difference Time Domain (FDTD) method is applied in the conductive regions, while the solution in the nonconductive regions is computed using the Boundary Element Method (BEM) ~\cite{Yioultsis}. In~\cite{Ausserhofer} an explicit time integration method and the discontinuous Galerkin finite element method (DG-FEM) are applied in conductive materials, while the finite element method based on continuous shape functions and an implicit time integration scheme are used in nonconductive domains~\cite{Ausserhofer}. In another recent approach, a similar DG-FEM explicit time stepping approach is used for an $H-\varPhi$ formulation of the magnetoquasistatic field problem \cite{Smajic}.

This work is based on an approach originally presented in~\cite{Clemens11}, where the magnetoquasistatic DAE based on an $\vec{A}^{*}-$field formulation is transformed into a finitely stiff ordinary differential equation (ODE) system by applying a generalized Schur complement. 

The structure of this paper is as follows: Section \ref{sec:Formulation} introduces the mathematical formulation of the eddy current problem and the transformation to an ordinary differential equation. In Section \ref{sec:mrhs} the time stepping and the resulting multiple right-hand side problem are discussed. Here, also the use of the subspace projection extrapolation method and of the proper orthogonal decomposition method as multiple right-hand side techniques is described. In Section \ref{sec:Validation} the simulation results for validating the presented approach and the effect of subspace projection extrapolation method and of the proper orthogonal decomposition method on a nonlinear test problem are presented. The main results of this paper are summarized in Section \ref{sec:conclusion}.

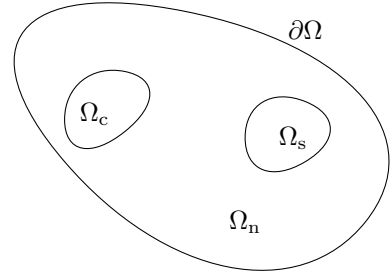
\begin{figure}
	\centering
	\begin{tikzpicture}[scale=0.8]
	\draw [black] plot [smooth cycle, tension=2] coordinates {(1,2) (3,4.5) (6,1)};
	\draw [black] plot [smooth cycle, tension=2] coordinates {(2,2.5) (2,3.5) (1,2.7)};
	\draw [black] plot [smooth cycle, tension=2] coordinates {(5,2) (5,3) (4,2.4)};
	\node at (4,1) {$\Omega_\mathrm{n}$};
	\node at (1.5,2.8) {$\Omega_\mathrm{c}$};
	\node at (4.8,2.4) {$\Omega_\mathrm{s}$};
	\node at (5,4.2) {$\partial\Omega$};
	\end{tikzpicture}
	\vspace{-0.5cm}
	\caption{\label{fig:kartoffel}Computational domain $\Omega$ split into three regions: conductive and nonlinearly permeable ($\Omega_\mathrm{c}$), nonconductive with constant permeability ($\Omega_\mathrm{n}$) and nonconductive with excitation ($\Omega_\mathrm{s}$).}
\end{figure}

\section{Mathematical Formulation}
\label{sec:Formulation}
The eddy current problem in the $\vec{A}^{*}-$formulation is given by the partial differential equation
\begin{equation}
\label{eq:pde}
\kappa\dfrac{\partial{\vec{\mathrm{A}}}}{\partial \mathrm{t}}+\nabla\times\left(\nu\left(\nabla\times\vec{\mathrm{A}}\right)\nabla\times\vec{\mathrm{A}}\right)=\vec{\mathrm{J}}_{\mathrm{S}},
\end{equation} where $\kappa$ is the electrical conductivity, $\vec{\mathrm{A}}$ is the time-dependent magnetic vector potential, $\nu$ is the reluctivity that can be nonlinear in ferromagnetic materials and  $\vec{\mathrm{J}}_{\mathrm{S}}=\vec{\mathrm{X}}_{\mathrm{S}}i_{\mathrm{S}}(t)$, where $i_{\mathrm{S}}(t)$  is the time-dependent source current and $\vec{\mathrm{X}}_{\mathrm{S}}$ distributes the current density spatially. Furthermore, initial values and boundary conditions are needed.
The weak formulation of (\ref{eq:pde}) leads to the variational problem: find $\vec{A}$
\begin{align*}
\int_\Omega \vec{w}&\cdot\kappa\frac{\partial\vec{A}}{\partial t}\;\mathrm{d}\Omega
\;+\\
&\int_\Omega \nabla\times\vec{w}\cdot\nu\left(\nabla\times\vec{A}\right)\nabla\times\vec{A}\;\mathrm{d}\Omega
=
\int_\Omega \vec{w}\cdot\vec{J}_{\mathrm{s}}\;\mathrm{d}\Omega
\end{align*}
for all $\vec{w}\in H_0(\mathbf{curl},\Omega)$ where we denote the spatial domain with $\Omega$ and assume Dirichlet conditions at the boundary $\partial\Omega$, see Fig.~\ref{fig:kartoffel}. Discretization and choosing test and ansatz functions from the same space according to Ritz-Galerkin, i.e., 
\begin{align}
\vec{A}(\vec{r},t)\approx\sum_{i=0}^{N_\mathrm{dof}} \vec{w}_i(\vec{r}) a_i(t)
\end{align}
leads to a spatially discretized symmetric equation system in time domain. Separation of the degrees of freedom (dofs) into two vectors $\mathbf{a}_\mathrm{c}$ storing the dofs allocated in conducting regions (if $\vec{r}\in\Omega_\mathrm{c}$) and $\mathbf{a}_\mathrm{n}$ holding the dofs allocated in nonconducting regions (if $\vec{r}\in\Omega_\mathrm{n}\cup\Omega_\mathrm{s}$) yields the DAE system 
\begin{equation}
\label{eq:DAE}
\begin{bmatrix}
\mathbf{M}_{\mathrm{c}} & 0 \\
0 & 0
\end{bmatrix} \dfrac{\mathrm{d}}{\mathrm{d}t} \begin{bmatrix}
\mathbf{a}_{\mathrm{c}}\\
\mathbf{a}_{\mathrm{n}}
\end{bmatrix} + \begin{bmatrix}
\mathbf{K}_{\mathrm{c}}(\mathbf{a}_{\mathrm{c}}) & \mathbf{K}_{\mathrm{cn}}\\
\mathbf{K}^{\top}_{\mathrm{cn}} & \mathbf{K}_{\mathrm{n}}
\end{bmatrix} \begin{bmatrix}
\mathbf{a}_{\mathrm{c}}\\
\mathbf{a}_{\mathrm{n}}
\end{bmatrix} = \begin{bmatrix}
0\\
\mathbf{j}_{\mathrm{Sn}} 
\end{bmatrix},
\end{equation} where $\mathbf{M}_{\mathrm{c}}$ is the conductivity matrix, $\mathbf{K}_{\mathrm{c}}$ is the nonlinear curl-curl reluctivity matrix in conducting regions,  $\mathbf{K}_{\mathrm{n}}$ is the typically constant curl-curl matrix in nonconducting regions, $\mathbf{K}_{\mathrm{cn}}$ is a coupling matrix, and $\mathbf{j}_{\mathrm{Sn}}$ is the source current typically defined in the nonconducting domain only. The conductivity matrix in (\ref{eq:DAE}) is not invertible and therefore the problems consists of differential-algebraic equations (DAEs). The numerical solution of these systems is more difficult than in the case of ordinary differential equations (ODEs). The level of difficulty is measured by the DAE index, which can be roughly interpreted as the number of differentiations needed to obtain an ODE from the DAE~\cite{Hairer}. System (\ref{eq:DAE}) is essentially an index-1 DAE with the speciality that the algebraic constraint, i.e., the second equation in (\ref{eq:DAE}), is formally not uniquely solvable for $\mathbf{a}_\mathrm{n}$ without defining a gauge condition due to the nullspace of the discrete curl-curl operator  $\mathbf{K}_{\mathrm{n}}$. However, it is well known that many iterative solvers have a weak gauging property, e.g. \cite{Clemens99}, such that a formal regularization can be avoided. 

Relying on this weak gauging property, the generalized Schur complement 
\begin{equation}
\label{eq:SC}
\mathbf{K}_{\mathrm{S}}(\mathbf{a}_{\mathrm{c}}):=\mathbf{K}_{\mathrm{c}}(\mathbf{a}_{\mathrm{c}})-\mathbf{K}_{\mathrm{cn}}\mathbf{K}^{+}_{\mathrm{n}}\mathbf{K}^{\top}_{\mathrm{cn}},
\end{equation} where $\mathbf{K}^{+}_{\mathrm{n}}$ represents a pseudo-inverse of $\mathbf{K}_{\mathrm{n}}$ in matrix form, is applied to (\ref{eq:DAE}) and transforms the DAE into
\begin{subequations}
	\begin{eqnarray}
	\mathbf{M}_{\mathrm{c}}\dfrac{\mathrm{d}}{\mathrm{d}t}\mathbf{a}_{\mathrm{c}}+\mathbf{K}_{\mathrm{S}}(\mathbf{a}_{\mathrm{c}})\mathbf{a}_{\mathrm{c}} &=& -\mathbf{K}_{\mathrm{cn}}\mathbf{K}^{\mathrm{+}}_{\mathrm{n}}\mathbf{j}_{\mathrm{s,n}}, \label{eq:ODE}\\
	\mathbf{a}_{\mathrm{n}} &=& \mathbf{K}^{+}_{\mathrm{n}}\mathbf{j}_{\mathrm{s,n}}-\mathbf{K}^{+}_{\mathrm{n}}\mathbf{K}^{\top}_{\mathrm{cn}}\mathbf{a}_{\mathrm{c}}. \label{eq:a_n}
	\end{eqnarray}
\end{subequations}

A regularization of $\mathbf{K}_{\mathrm{n}}$ by a grad-div or tree/cotree gauging can be used alternatively~\cite{Clemens11,Schoeps}. Here, the pseudo-inverse is evaluated using the preconditioned conjugate gradient method (PCG)~\cite{Dutine}. The finitely stiff ODE (\ref{eq:ODE}) can be integrated explicitly in time, e.g. by using the explicit Euler method. Using this time integration method, the expressions
\begin{align}
\mathbf{a}^{m}_{\mathrm{c}}
&= \mathbf{a}^{m-1}_{\mathrm{c}}\!+\!\Delta t\mathbf{M}^{-1}_{\mathrm{c}}\left[\mathbf{K}_{\mathrm{cn}}\mathbf{K}^{+}_{\mathrm{n}}\mathbf{j}^{m}_{\mathrm{s,n}}\!-\!\mathbf{K}_{\mathrm{S}}(\mathbf{a}^{m-1}_{\mathrm{c}})\mathbf{a}^{m-1}_{\mathrm{c}}\right]\!,\! \label{eq:acm}\\
\mathbf{a}^{m}_{\mathrm{n}} &= \mathbf{K}^{+}_{\mathrm{n}}\mathbf{j}^{m}_{\mathrm{s,n}}-\mathbf{K}^{+}_{\mathrm{n}}\mathbf{K}^{\top}_{\mathrm{cn}}\mathbf{a}^{m}_{\mathrm{c}} \label{eq:anm}
\end{align}
are computed in the $m$-th time step, where $\Delta t$ is the time step size. The Courant-Friedrich-Levy (CFL) criterion determines the maximum stable time step size of explicit time integration methods~\cite{Hairer}. For the explicit Euler method
\begin{equation}
\label{eq:CFL}
\Delta t\leq\dfrac{2}{\lambda_{\mathrm{max}}\left(\mathbf{M}^{-1}_{\mathrm{c}}\mathbf{K}_{\mathrm{S}}\left(\mathbf{a}_{\mathrm{c}}\right)\right)}
\end{equation} is an estimation for the maximum stable time step size, where $\lambda_{\mathrm{max}}$ is the maximum eigenvalue~\cite{Schoeps}. The maximum eigenvalue can be estimated using the power method~\cite{Golub}.

\section{Multiple Right-Hand Side Problem}
\label{sec:mrhs}

As the matrix  $\mathbf{K}_{\mathrm{n}}$ remains constant within each explicit time step, the repeated evaluation of a pseudo-inverse $\mathbf{K}^{+}_{\mathrm{n}}$ in (\ref{eq:acm}), (\ref{eq:anm}) forms a multiple right-hand side (mrhs) problem of the form
\begin{equation}
\label{eq:mrhs}
\mathbf{K}_{\mathrm{n}}\mathbf{a}_{\mathrm{p}}=\mathbf{j}_{\mathrm{p}}\Leftrightarrow\mathbf{a}_{\mathrm{p}}=\mathbf{K}^{+}_{\mathrm{n}}\mathbf{j}_{\mathrm{p}}.
\end{equation}Here, $\mathbf{j}_{\mathrm{p}}$ represents one of the right-hand side vectors $\mathbf{j}^{m}_{\mathrm{s,n}}$,$\,\mathbf{K}^{\top}_{\mathrm{cn}}\mathbf{a}^{m}_{\mathrm{c}}$, and $\mathbf{K}^{\top}_{\mathrm{cn}}\mathbf{a}^{m-1}_{\mathrm{c}}$. Instead of computing the matrix $\mathbf{K}^{+}_{\mathrm{n}}$ explicitly, a vector $\mathbf{a}_\mathrm{p}$ is computed according to (\ref{eq:mrhs}) using the preconditioned conjugate gradient (PCG) method~\cite{Dutine}.

Improved start vectors for the PCG method can be obtained by the subspace projection extrapolation (SPE) method or the proper orthogonal decomposition (POD) method.
In the SPE method, the linearly independent column vectors of a matrix $\mathbf{U}_{\mathrm{SPE}}$ are formed by a linear combination of an orthonormalized basis of the subspace spanned by solutions $\mathbf{a}_{\mathrm{p}}$ from  previous time steps. The modified Gram-Schmidt method is used for this orthonormalization procedure~\cite{Trefethen}. The improved start vector $\mathbf{x}_{\mathrm{0,SPE}}$ is then computed by~\cite{Clemens04}
\begin{equation}
\label{eq:SPE_x0}
\mathbf{x}_{\mathrm{0,SPE}}:=\mathbf{U}_{\mathrm{SPE}}\left(\mathbf{U}^{\top}_{\mathrm{SPE}}\mathbf{K}_{\mathrm{n}}\mathbf{U}_{\mathrm{SPE}}\right)^{-1}\mathbf{U}^{\top}_{\mathrm{SPE}}\mathbf{K}^{\top}_{\mathrm{cn}}\mathbf{j}_{\mathrm{p}}.
\end{equation} 

As only the last column vector in the matrix $\mathbf{U}_{\mathrm{SPE}}$ changes in every time step, all other matrix-column-vector products in computing $\mathbf{K}_\mathrm{n}\mathbf{U}_{\mathrm{SPE}}$ in (\ref{eq:SPE_x0}) are reused from previous time steps in a modification of the procedure in~\cite{Clemens04} referred to as the "Cascaded SPE" (CSPE)~\cite{Dutine}.
 
When using the POD method for the PCG start vector generation, $N_{\mathrm{POD}}$ solution vectors from previous time steps form the column vectors of a snapshot matrix $\mathbf{X}$ which is decomposed into
\begin{equation}
\label{eq:SVD}
\mathbf{X}=\mathbf{U}\mathbf{\Sigma}\mathbf{V}^{\top}
\end{equation}
using the singular value decomposition (SVD)~\cite{Henneron14, Henneron15, Sato}. Here, $\mathbf{U}$ and $\mathbf{V}$ are orthonormal matrices and $\mathbf{\Sigma}$ is a diagonal matrix of the singular values ordered by magnitude $\left(\sigma_{\mathrm{i}}\geq\sigma_{\mathrm{j}}\,\mathrm{for}\,i<j\right)$. The index $k$ is chosen such that the information of the largest singular values is kept
 \begin{equation}
 \dfrac{\sigma_{\mathrm{k}}}{\sigma_{1}}\leq\varepsilon_\mathrm{POD}. \label{eq:relSV2}
 \end{equation} The threshold value $\varepsilon_\mathrm{POD}$ is here chosen as $\varepsilon_\mathrm{POD}:=10^{-4}$. A measure how much information is kept can be computed by
 the relative information criterion~\cite{Daniel}
 \begin{equation}
 \label{eq:keepInfo}
\dfrac{\sum\limits_{i=1}^{k}\sigma_{i}}{\sum\limits_{i=1}^{N_{\mathrm{POD}}}\sigma_{i}}\overset{!}{\approx}1.
 \end{equation}
Defining $\mathbf{U}_{\mathrm{POD}}=\left[\mathbf{U}_{\mathrm{:,1}},...\,,\mathbf{U}_{\mathrm{:,k}}\right]$ as the first $k$ columns of $\mathbf{U}$ allows to compute an improved start vector $\mathbf{x}_{\mathrm{0,POD}}$ by 
 \begin{equation}
 \label{eq:POD_x0}
 \mathbf{x}_\mathrm{0,POD}:=\mathbf{U}_{\mathrm{POD}}\left[\mathbf{U}^{\top}_{\mathrm{POD}}\mathbf{K}_{\mathrm{n}}\mathbf{U}_{\mathrm{POD}}\right]^{-1}\mathbf{U}^{\top}_{\mathrm{POD}}\mathbf{K}^{\top}_{\mathrm{cn}}\mathbf{j}_{\mathrm{p}}.
 \end{equation}

The repeated evaluation of $\mathbf{M}^{-1}_{\mathrm{c}}$ in (\ref{eq:acm}) also forms a mrhs problem, and both the POD and the CSPE method can be used for computing improved start vectors
for the PCG method. In the case of small matrix dimensions of the regular matrix $\mathbf{M}_{\mathrm{c}}$, the inverse can also be computed directly using GPU-acceleration.

\section{Numerical Validation}
\label{sec:Validation}
\begin{figure}[!t]
	\centering
	\includegraphics[width=3.6in]{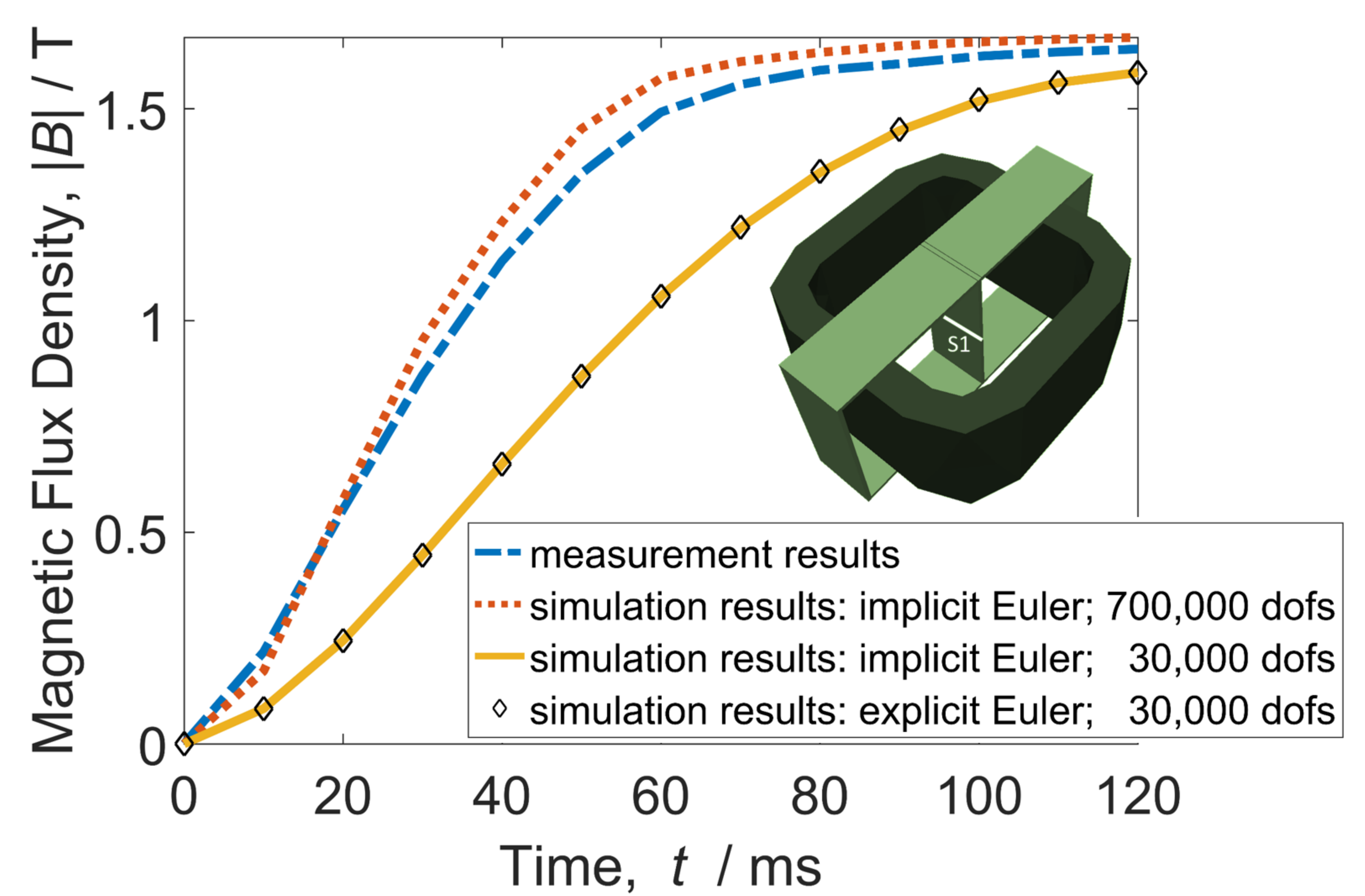}
    \caption{Comparison of results for the average magnetic flux density evaluated at position S1 and model geometry as inset.}
	\label{fig:fig_1}
\end{figure}

The ferromagnetic TEAM 10 benchmark problem is used for numerical validation of the presented explicit time integration scheme for magnetoquasistatic fields~\cite{Nakata}. The domain consists of two square-bracket-shaped steel plates opposite of each other and a rectangular steel plate between them, resulting in two 0.5 mm wide air gaps. The model geometry is shown in Fig. \ref{fig:fig_1}. The position where the magnetic field is evaluated is marked as S1. The excitation current $i_{\mathrm{S}}=(1-\exp(-t/\tau))$, where $\tau=0.5\,\mathrm{s}$, is applied for a time interval of 120 ms starting at $t=0\,\mathrm{s}$~\cite{Nakata}. The resulting magnetic flux density is computed for this time interval.

The finite element method (FEM) using 1st order edge elements is used for the spatial discretization~\cite{Kameari}. All simulations are computed on a workstation with an Intel Xeon E5 processor and an NVIDIA TESLA K80 GPU. The conjugate gradient method is preconditioned by an algebraic multigrid method \cite{AMG}. The matrix $\mathbf{M}_{\mathrm{c}}$ is inverted using the Magma-library and GPU-acceleration~\cite{Magma}.

A fine mesh resulting in about 700,000 dofs and the implicit Euler method are used to validate the simulation code. A good agreement between the measured results published in~\cite{Nakata} and the simulation of this fine spatial discretization is shown in Fig. \ref{fig:fig_1}. The required simulation time of this simulation using the implicit Euler method is 5.38 days using an in-house implicit time integration magnetoquasistatic code. 

\begin{figure}[!t]
	\hspace*{-0.5cm}
	\includegraphics[width=3.6in]{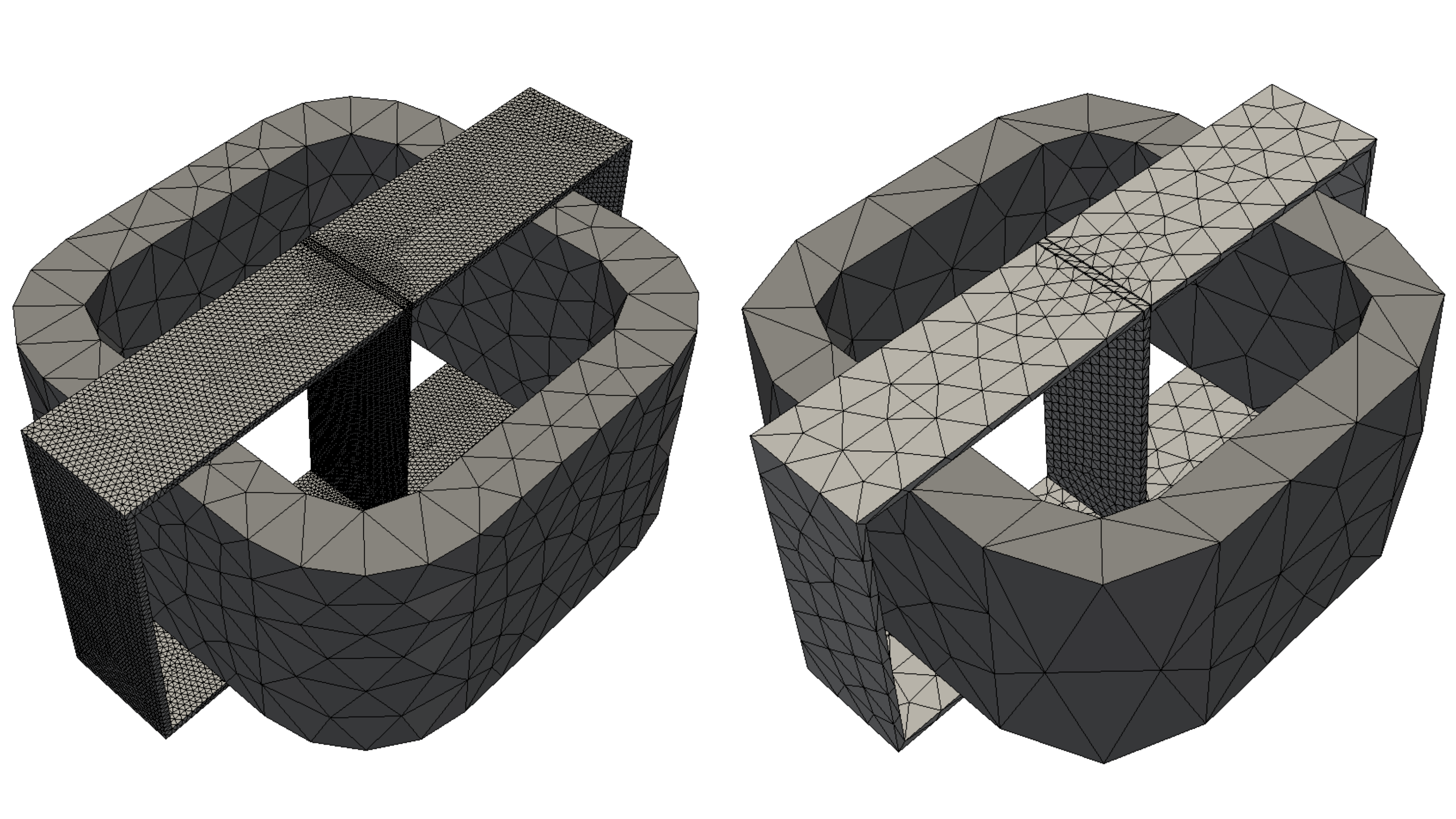}
	\caption{Meshes resulting in about 700,000 dofs (left) and in about 30,000 dofs (right).}
	\label{fig:meshes}
\end{figure}

\begin{figure}[!t]
	\hspace*{-0.5cm}
	\includegraphics[width=3.6in]{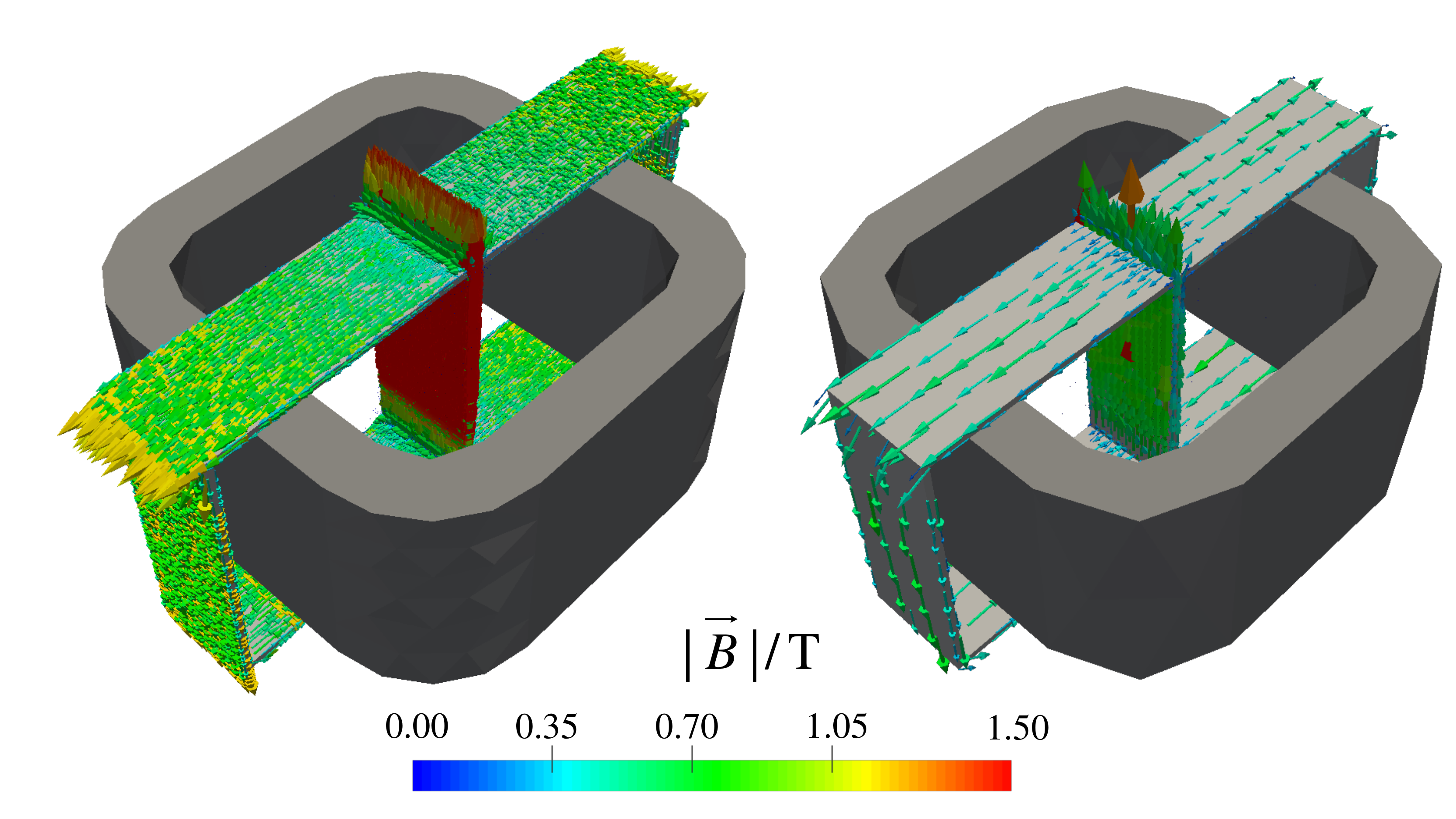}
	\caption{Field plots of the magnetic flux density $\vec{B}$ for the spatial discretization with about 700,000 dofs (left) and with about 30,000 dofs (right).}
	\label{fig:fieldPlots}
\end{figure}

For benchmarking the proposed mrhs techniques for the (semi-)explicit time integration scheme, a model with a coarse spatial discretization yielding about 30,000 dofs and the explicit Euler method is used. For this spatial discretization, the resulting maximum stable time step size according to (\ref{eq:CFL}) is $\Delta t_{\mathrm{CFL}}=1.2\,\mu s$. Both meshes are presented in Fig. \ref{fig:meshes}
The results for the average magnetic flux density are compared with the results obtained using the same discretization in space and the implicit Euler method for time integration and show good agreement, depicted in Fig. \ref{fig:fig_1}. The resulting field plots for both spatial discretizations are shown in Fig. \ref{fig:fieldPlots}. The simulation time for the implicit time integration method is still 2.58 h.

\begin{figure}[!t]
	\centering
	\includegraphics[width=3.4in]{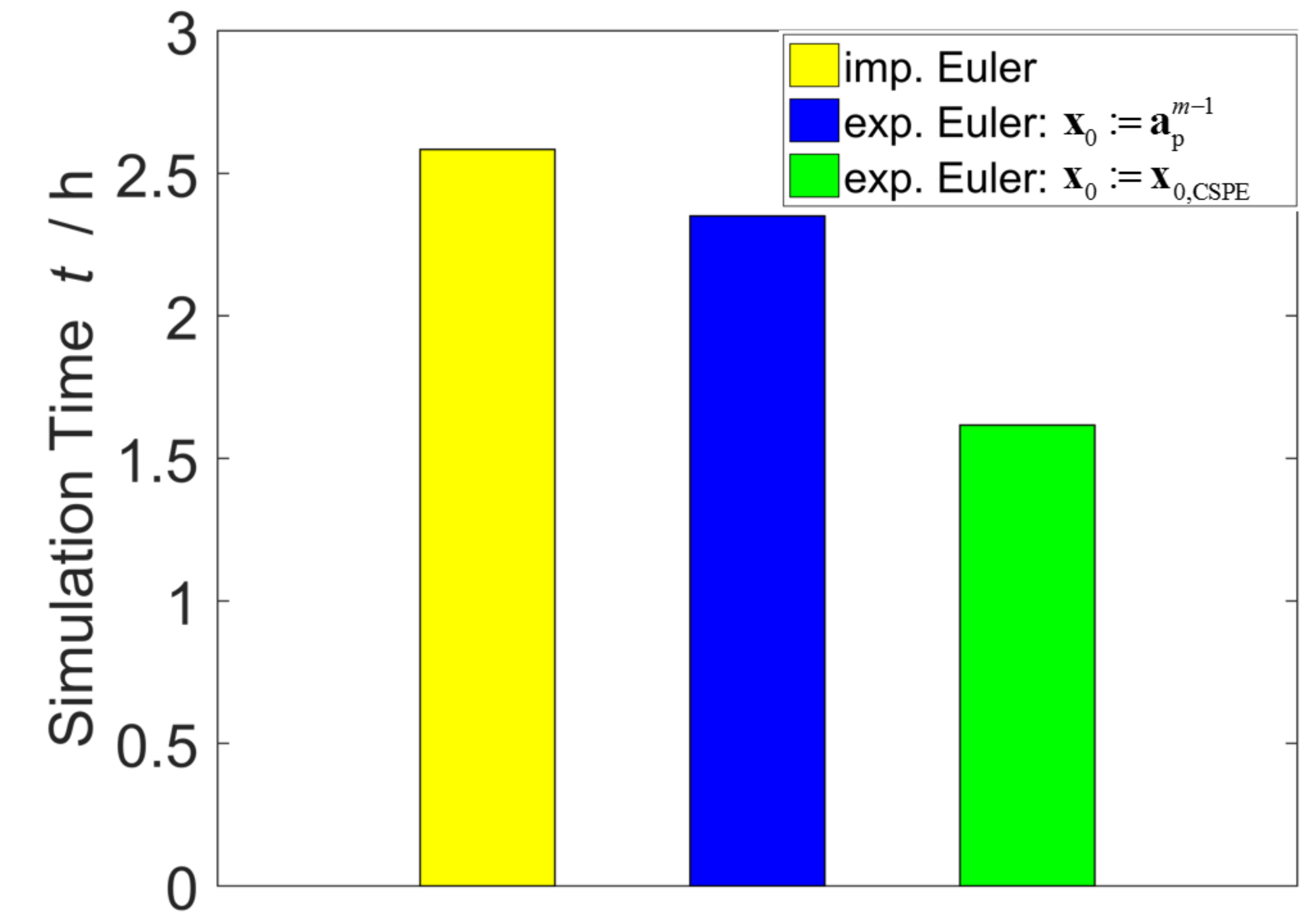}
	\caption{Comparison of simulation times.}
	\label{fig:simTimes}
\end{figure}

The effect of computing improved start vectors using POD or CSPE on the average number of PCG iterations and on the simulation time is compared to using the solution from the previous time step $\mathbf{a}^{m-1}_{\mathrm{p}}$ as start vector for the PCG method. An overview is presented in Table \ref{table1} and shows that both the CSPE and the POD start vector generation methods significantly reduce the number of PCG iterations. When using CSPE the number of column vectors in the operator $\mathbf{U}_{\mathrm{SPE}}$ in (\ref{eq:SPE_x0}) is increased during the simulation to improve the spectral information content of $\mathbf{U}_{\mathrm{SPE}}$. This number remains below 20. Thus, only small systems have to be solved for the inversion in (\ref{eq:SPE_x0}) and the effort to perform all computations of the CSPE method is low. This is also confirmed by the simulation time which is shortest when using CSPE. The simulation time resulting from using explicit time integration and CSPE for start vector generation is  $63\,\%$ of the simulation time of the implicit reference simulation. A bar plot showing the reduced simulation time by using the explicit Euler scheme and CSPE compared to using the standard formulation and the implicit Euler method for time integration is depicted in Fig. \ref{fig:simTimes}.

In case of the POD, the amount of information kept according to (\ref{eq:keepInfo}) is $>0.99$ during the entire simulation. However, the computational effort for performing the SVD and the computations in (\ref{eq:POD_x0}) is higher than the effort for CSPE. Although the number of PCG iterations is further decreased, the simulation time resulting from using POD for start vector generation is higher than when using  $\mathbf{a}^{m-1}_{\mathrm{p}}$ as start vector for the PCG method due to the costs of the repeated SVD.

\begin{table}[!t]
\renewcommand{\arraystretch}{1.3}
\caption{Simulation Time and Average Number of PCG Iterations Using Different Start Vectors $\mathbf{x}_{\mathrm{0}}$}
\label{table1}
\centering
\begin{tabular}{|c||c||c|}
\hline
 Start vector & Avg. Number of PCG Iterations & Simulation Time\\
\hline
$\mathbf{x}_{\mathrm{0}}:=\mathbf{a}^{m-1}_{\mathrm{p}}$ & 3.16 & \, 2.35 h\\
\hline
$\mathbf{x}_{\mathrm{0}}:=\mathbf{x}_{\mathrm{0,POD}}$ & 2.18 & 17.35 h\\
\hline
$\mathbf{x}_{\mathrm{0}}:=\mathbf{x}_{\mathrm{0,CSPE}}$ & 1.02 & \, 1.62 h\\
\hline
\end{tabular}
\end{table}

\section{Conclusion}
\label{sec:conclusion}
The magnetic vector potential formulation of eddy current problems was transformed into an ODE system of finite stiffness using a generalized Schur complement.
The resulting ODE system was integrated in time using the explicit Euler method. A pseudo-inverse of the singular curl-curl matrix in nonconducting material was evaluated using the PCG method. 

Improved start vectors for the PCG method were calculated using the POD and the CSPE method. Although both reduce the number of PCG iterations needed, the computational effort of the CSPE is significantly lower than for the POD method. Reducing the computational effort of the POD, e.g. by accelerating the computation of the SVD is subject to further investigations.
Using the CSPE method, the overall simulation time was reduced by $37\,\%$ compared to the simulation time of the implicit reference simulation.

\section*{Acknowledgment}

This work was supported by the Deutsche Forschungsgemeinschaft (DFG) under grant numbers CL143/11-1 and SCHO1562/1-1. The third author is supported by the “Excellence Initiative” of the German Federal and State Governments and The Graduate School of CE at TU Darmstadt.

\end{document}